# Ultra-high energy cosmic rays from supermassive black holes: particle flux on the Earth and extragalactic diffuse emission

A V Uryson

**Lebedev Physical Institute of RAS Moscow 119991**

E-mail: uryson@sci.lebedev.ru

**Abstract.** Cosmic rays accelerated to ultra-high energies ($E > 4 \cdot 10^{19}$ eV) in electric fields in accretion discs around supermassive black holes are discussed. Particle injection spectra are assumed to be harder than those formed in acceleration at shock fronts. It turned out that cosmic rays with injection spectra supposed contribute a little in the particle flux detected with ground-based arrays. But in the extragalactic space particles generate a noticeable flux of the diffuse gamma-ray emission compared with the data obtained with Fermi LAT instrument (onboard the Fermi space observatory). The intensity of neutrinos generated by cosmic rays propagating in the extragalactic space is also derived. The model intensity of cascade neutrinos is much lower than that of astrophysics neutrino. It is concluded that possibly supermassive black holes are cosmic ray sources which contribute slightly to the particle flux detected on the Earth, but these sources should be taken into account analyzing components of extragalactic diffuse gamma-ray emission.

## 1. Introduction

Sources of cosmic rays (CRs) at ultra-high-energies (UHE), $E > 4 \cdot 10^{19}$ eV, are still unknown. Apparently UHECRs are of extragalactic origin being emitted by active galactic nuclei.

UHECRs are investigated with giant ground arrays Pierre Auger Observatory (PAO) and Telescope Array (TA). The data obtained are particle arrival directions, particle energies and mass composition (protons or atomic nuclei).

Assuming rectilinear CR propagation in space UHECR sources can be identified using particle arrival directions. However this approach was unsuccessful mainly by two reasons. First, CRs are apparently deflected by extragalactic magnetic fields. Second, error-boxes in particle arrivals are of $\sim 1^0$ so they contain many objects among which it is difficult to identify the particle source.

In extragalactic space UHECRs interact with background emission which leads to the lack of particles at energies $E > 10^{20}$ eV if CRs fly distances more than ~100 Mpc (GZK-suppression: [1, 2]). Energy spectra obtained both with PAO and TA are suppressed. But the spectra disagree in shape and that is interpreted as different mass composition of CR detected: according to the PAO data UHECRs are protons, while according to the TA data they are nuclei.

Apart from GZK-suppression UHECR interacting with background emission produces electromagnetic cascades in extragalactic space, in which gamma-ray emission is generated [3, 4]. It is the component of the diffuse extragalactic emission measured with Fermi LAT (Large Area Telescope on board cosmic observatory Fermi) [5].

Thus CRs are now investigated applying not only CR data (energetic spectra etc.) but also data on cascade gamma-ray emission. Therefore CR models should meet now two criteria. First, it is required that the calculated UHECR energy spectra fit the spectrum measured. Second, the model intensity of the cascade gamma-ray emission must be lower than the measured intensity of the extragalactic diffuse emission minus the contribution of individual unresolved gamma-ray sources. Based on this scheme, the CR data were analyzed in e.g. [6-9].

In [6] the model [10] of CR acceleration by supermassive black holes is investigated. Papers [7, 8] are concerned with UHECR mass composition (only protons or protons and nuclei). Paper [9] is dedicated to dark matter models, because the decays of dark matter particles contribute to the diffuse gamma-ray emission, so data about all components of diffuse emission are required to analyze dark matter. In [6] an ensemble of supermassive black holes (SMBH) is assumed as a set of possible UHECR sources, with monoenergetic injection spectra depending on SMBH mass. It is concluded that the model fits well the measured UHECR spectrum being consistent with the Fermi LAT data on diffuse gamma-ray emission. In papers [7-9] the model injection spectra along with the cosmological evolution of possible UHECR sources are varied to provide the best fit with the UHECR spectrum. Then models are chosen where cascade gamma-ray intensity is lower than the measured intensity of the extragalactic diffuse emission minus the contribution of individual unresolved gamma-ray sources. In these papers model injection spectra are taken to be exponential with the index (depending on the source cosmological evolution) $|\alpha|\approx2.2-2.6$. These values of the exponent are typical for particle acceleration on shock fronts, which can occur, for example, in active galactic nuclei jets.

In this paper we discuss UHECRs accelerated in electric fields in accretion discs around SMBHs [11], assuming the injection spectra to be harder than abovementioned ones and including equiprobable generation of particles at any energy in between $10^{19} - 10^{21}$ eV. In the latter case the spectral index is $\alpha=0$. We obtain that UHECR sources having these injection spectra contribute negligibly to the CR flux detected by ground-based arrays. Calculated spectra of particles reaching the Earth differs significantly in shape from the CR spectrum measured. However, despite the insignificant intensity of particles near the Earth, CRs from these sources can produce a noticeable diffuse gamma-ray flux in intergalactic space compared to the Fermi LAT data. This should be taken into account while analyzing CR and dark matter models, because both the UHECR cascade emission and the decay of dark matter particles contribute to the diffuse emission. We also conclude that gamma-ray emission data possibly can be applied to study UHECR acceleration in SMBHs.

Neutrinos are also produced via UHECR interaction with background emission and they contribute to the astrophysical neutrino flux. Data on neutrino flux is currently obtained with IceCube array. Neutrinos produced via UHECR interaction in extragalactic space are calculated in e.g. [7-9] and [12, 13]. In these papers it is shown that neutrino data restrict UHECR models lesser than gamma-ray emission data. In this paper we also calculate neutrino spectra and obtain that model neutrino intensity is much lower than the measured one.

The computations were performed with the TransportCR code [14].

## 2. The model

Electromagnetic cascades result mainly from the following reactions [3, 4]. In intergalactic space UHECRs interact with microwave and radio emissions $p+\gamma_{rel} \rightarrow p+\pi^0$, $p+\gamma_{rel} \rightarrow n+\pi^+$. Pions decay $\pi^0 \rightarrow \gamma+\gamma$, $\pi^+ \rightarrow \mu^+ +\nu_\mu$ giving rise to gamma-quanta and muons, and muons decay $\mu^+ \rightarrow e^+ +\nu_e+\bar{\nu}_\mu$ giving rise to positrons and neutrinos. Gamma-quanta and positrons generate electromagnetic cascades in the reactions with microwave emission and extragalactic background light $\gamma+\gamma_b \rightarrow e^+ + e^-$ (pair production) and $e+ \gamma_b \rightarrow e'+ \gamma'$ (inverse Compton effect).

The assumptions adopted in the model refer to three items: the injection spectra and evolution of UHECR sources, extragalactic background emissions, and extragalactic magnetic fields.

We assume that the UHECR sources are pointlike. These are AGNs where charged particles are accelerated to UHE by electric fields in accretion discs around supermassive black holes [11]. Due to this acceleration mechanism we assume the CR injection spectrum to be harder than that when particles are accelerated at shock fronts. We assume the spectral index to be $|\alpha| \leq 2.2$, including equiprobable generation of particles at any energy in between $10^{19} - 10^{21}$ eV, when $\alpha=0$.
We also assume that UHECRs consist of protons.
We assume that UHECR sources under consideration are located at distances with red shifts $z \geq 0.05$. Source evolution apparently affect the CR spectrum near the Earth (see, e.g. [8, 15]). SMBH evolution is associated with the evolution of their states (see, e.g., [16]). It is unclear and therefore we consider the evolution scenario [17] of powerful active galactic nuclei - Blue Lacertae objects that was discussed in [7, 14].
The extragalactic background emissions were treated as follows.
The cosmic microwave background radiation has Planck energy distribution with the mean value $\varepsilon_r = 6.7 \cdot 10^{-4}$ eV. The mean photon density is $n_r = 400$ cm$^{-3}$.
The extragalactic background light characteristics were taken from [18].
To describe the background radio emission, we used the model of the luminosity evolution for radio galaxies [19].
The magnetic field in intergalactic space is apparently nonuniform [20, 21]: there are regions where the magnetic field is $1 \cdot 10^{-17}$ G < B < $3 \cdot 10^{-14}$ G, and filamentary areas in which the magnetic field is stronger: $B \approx 10^{-9} - 10^{-8}$ G. In these fields, the cascade electrons lose their energy through synchrotron radiation insignificantly [22]. However inside galactic clusters the magnetic field possibly is higher, up to B$\sim 10^{-6}$ G [23]. This field disturbs cascades due to electron synchrotron emission. We suppose that regions where extragalactic fields are of $\sim 10^{-6}$ G occupy a small part of extragalactic space and neglect them. So we assume that extragalactic magnetic fields do not break cascades.

## 3. Results

The calculated UHECR energy spectra along with the PAO spectrum are shown in figure 1. The model spectra are normalized to the PAO spectrum at the energy of $10^{19.5}$ eV ($3.16 \cdot 10^{19}$ eV). The model CR spectra are lower than the PAO spectrum by several orders of magnitude. In addition the calculated spectra differ greatly from the PAO spectrum in shape.
We compare the model spectra with the PAO one, although UHECRs are assumed to be protons in the model, while according to the PAO data, these are nuclei. The reason for such comparison is as follows. At energies $E \leq 10^{19.5}$ eV the difference between TA and PAO spectra is small: 20–30%. At $E \geq 5 \cdot 10^{19}$ eV it is large: the PAO spectrum is lower than the TA one by a factor of 8–9 [24]. Therefore, the model CR spectra are lower than both PAO and TA spectra by several orders of magnitude.
We will now proceed to the intensity of gamma-ray emission that UHECRs initiate in extragalactic space.
Spectra of the cascade gamma-ray emission are analyzed in detail in [23, 25] where it is found that the shape of the cascade spectrum is virtually independent on the initial CR spectrum. Thence here we analyze only the intensity of cascade gamma-ray emission without discussing the spectra. Specifically, we compare model integral intensity of cascade gamma-ray emission with Fermi LAT data in the range $E > 50$ GeV. The reason is that in this range the contribution of discrete unresolved gamma-ray sources is estimated in [26]. It is taken into account below when comparing the model cascade gamma-ray intensity with Fermi LAT result. (The similar comparison is done in [8].)
The integral intensity of the cascade gamma-ray emission $I_\gamma$ ($E > 50$ GeV) is found from the differential intensity, which is calculated with the TransportCR code. For the set of spectral index values $I_\gamma$ ($E > 50$ GeV) equals to:

$$\alpha=0: \quad I\gamma\ (E>50\ \text{GeV}) = 5.416 \cdot 10^{-10}\ (\text{cm}^{-2}\ \text{s}^{-1}\ \text{sr}^{-1}), \quad (1)$$
$$\alpha=-0.5: \quad I\gamma\ (E>\ \text{GeV}) = 4.123 \cdot 10^{-10}\ (\text{cm}^{-2}\ \text{s}^{-1}\ \text{sr}^{-1}), \quad (2)$$
$$\alpha=-1: \quad I\gamma\ (E>50\ \text{GeV}) = 3.056 \cdot 10^{-10}\ (\text{cm}^{-2}\ \text{s}^{-1}\ \text{sr}^{-1}), \quad (3)$$
$$\alpha=-1.8: \quad I\gamma\ (E>50\ \text{GeV}) = 2.015 \cdot 10^{-10}\ (\text{cm}^{-2}\ \text{s}^{-1}\ \text{sr}^{-1}), \quad (4)$$
$$\alpha=-2.2: \quad I\gamma\ (E>50\ \text{GeV}) = 2.376 \cdot 10^{-11}\ (\text{cm}^{-2}\ \text{s}^{-1}\ \text{sr}^{-1}). \quad (5)$$

Cascade neutrino intensity calculated in the range $E \approx 10^{11}$-$10^{15}$ eV is several orders lower than that measured by IceCube [27].

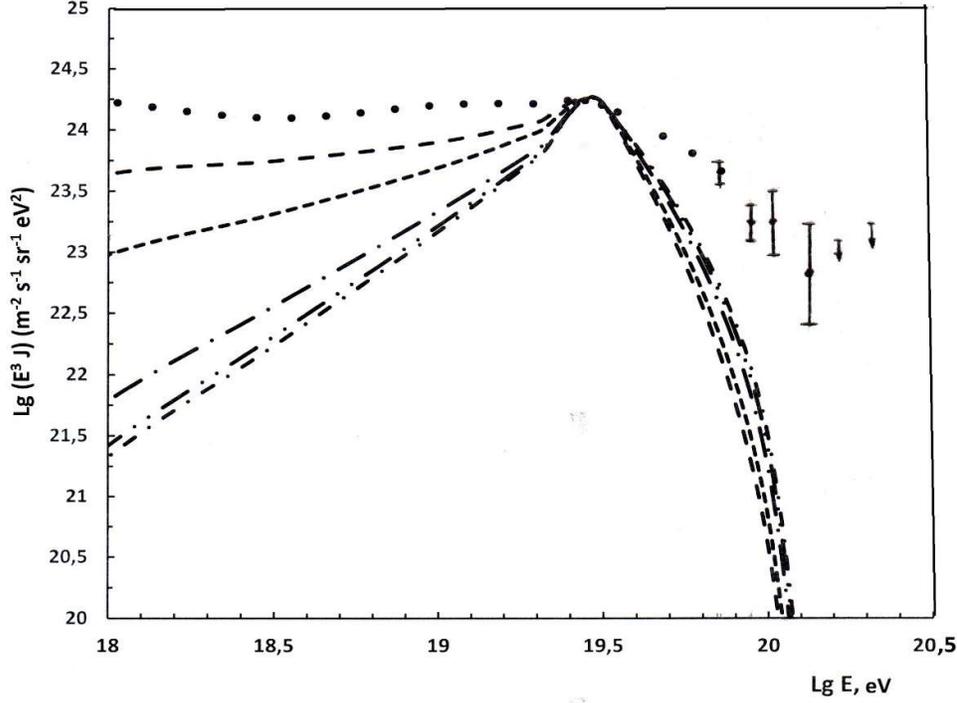

Fig.1. UHECR energy spectra: PAO data – points, and model spectra calculated with various spectral index values: long dashed line – α=2.2, dashed line - α=1.8, long dash dotted line - α=1, dash dotted line with two dots - α=0.5, short dash dotted line - α=0.

## 4. Discussion

The model intensity of cascade gamma-ray emission depends on the proton injection spectrum. There are two reasons for this [30]: the resonance in the energy dependence of the proton–relic photon ($p$–$\gamma_{rel}$) interaction cross section and the shape of the initial CR spectrum. The emitted proton interacts with the microwave background on its way from the source until its energy decreases to approximately $4 \cdot 10^{19}$ eV. The mean free path of the proton is then hundreds of Mpc, and the probability of its interaction with the background radiation becomes low. Protons emitted at energies of about $10^{20}$ eV and higher interacts with the background ~10 times on its way to the Galaxy, initiating approximately ten electromagnetic cascades in contrast with 1-2 cascades which are initiated by protons at energies of $5 \cdot 10^{19}$ eV. Thus the higher the proton energy the more effective is its energy transfer to the cascade and therefore in gamma-ray emission. As a result the proton energy transfer is more effective the harder the injection spectrum is. In addition, the cascade gamma-ray intensity depends on the distance to the source: distances of tens Mpc are too short for a cascade to develop.

Now we will compare the model intensity of the cascade gamma-ray emission with the Fermi LAT data. Extragalactic isotropic diffuse gamma-ray background measured by Fermi LAT [5] is:

$$\text{IGRB } (E>50 \text{ GeV}) = 1.325 \cdot 10^{-9} \text{ (cm}^{-2} \text{ s}^{-1} \text{ sr}^{-1}). \tag{6}$$

This value includes the emission from individual unresolved gamma-ray sources. Their contribution to the IGRB at energies $E>50$ GeV equals to 86 (−14,+16)% [28]. Subtracting from the IGRB the unresolved source contribution of 86%, we obtain

$$\text{IGRB }_{\text{without blazars}} (E>50 \text{ GeV}) = 1.855 \cdot 10^{-10} \text{ (cm}^{-2} \text{ s}^{-1} \text{ sr}^{-1}). \tag{7}$$

The cascade gamma-ray emission being the component of IGRB, its intensity should satisfy the condition

$$I\gamma \, (E>50 \text{ GeV}) < \text{IGRB }_{\text{without blazars}} (E>50 \text{ GeV}). \tag{8}$$

Unresolved gamma-ray source contribution to the IGRB to be 72% (based on the error -14% [28]) we obtain in place of (7):

IGRB $_{without\ blazars}$ ($E$>50 GeV) =3.71·10$^{-10}$ (cm$^{-2}$ s$^{-1}$ sr$^{-1}$). (9)

With this value the condition (8) is satisfied if the injection spectral index is α≤-1.

Three models of the galactic background - "A", "B", and "C" - are used in [5] when analyzing the data. Allowing for measurement errors along with the unresolved gamma-ray source contribution of 72% (rather than 86%), and the uncertainty of galactic background in the model "A", the band of IGRB $_{without\ blazars}$ ($E$>50 GeV) is obtained:

2.20·10$^{-10}$ (cm$^{-2}$ s$^{-1}$ sr$^{-1}$) ≤IGRB $_{without\ blazars}$ ($E$>50 GeV) ≤ 5.40·10$^{-10}$ (cm$^{-2}$ s$^{-1}$ sr$^{-1}$). (10)

The intensity of cascade gamma-ray emission $I\gamma$ ($E$>50 GeV) satisfy (10) at values of injection spectral index: α=0, -0.5, -1, -1.8, -2.2. (At α=0 $I\gamma$ ($E$>50 GeV) exceeds the upper limit (10) by 0.3%.) Source evolution also influences on cascade gamma-ray intensity: stronger evolution leads to increase of gamma-ray intensity [28].

Here we do not analyze the cascade gamma-ray emission spectrum, because its shape does not depend on the UHECR injection spectrum [23, 25].

At the energies $E$ >10$^{19}$ eV the intensity and the form of the cascade neutrino model spectrum depend on the injection CR spectrum. However this energy range is inaccessible for operating neutrino telescopes. So currently it is difficult to apply neutrino data for analyzing the model under consideration.

## 5. Conclusion

We analyzed UHECRs accelerated in electric fields in accretion discs surrounding SMBHs. Evolution of SMBHs was assumed to be the same as that of powerful active galactic nuclei – Blue Lacertae objects. Assuming that model injection spectra are exponential with the index α=0, -0.5, -1, -1.8, -2.2, that is harder than in acceleration on shock fronts, we calculate CR flux near the Earth. As a result of hard injection spectra in the model UHECRs contribute slightly to the particle flux detected by ground-based arrays. However in the model UHECRs produce a noticeable diffuse gamma-ray flux, in comparison with Fermi LAT data. It is important to allow for the possible contribution of these UHECRs analyzing UHECR sources, UHECR composition and dark matter decay. Possibly UHECR data can be applied when studying processes in SMBH accretion discs.

At the energies $E$ >10$^{19}$ eV the intensity and the form of the model cascade neutrino spectrum depend on the injection CR spectrum. However this energy range is inaccessible for current neutrino telescopes. The results obtained depend on the contribution of individual unresolved gamma-ray sources to the extragalactic diffuse emission. At present, it has been determined with an error of about 15% ranging from 72 to 100% [26]. This value can be refined using instruments with a better angular resolution than that of Fermi LAT. For example, the GAMMA-400 space instrument has an angular resolution of ∼0.01$^0$ at an energy of 100 GeV [29], while Fermi LAT angular resolution at the same energy is 0.05$^0$ –0.1$^0$ [30].

**Acknowledgments**
The author is grateful to O.E. Kalashev for the discussion of TransportCR code and N.P. Topchiev for the discussion of space based gamma-ray telescope characteristics.